\documentclass[sigconf,usgov]{acmart}
\usepackage{array}
\usepackage{color,soul}
\usepackage{balance}
\usepackage{xspace}
\usepackage{fancyvrb}
\usepackage{url}

\patchcmd{\quote}{\rightmargin}{\leftmargin 1.0em \rightmargin}{}{}

\newcommand{\omitt}[1]{}

\copyrightyear{2020} 
\acmYear{2020} 
\setcopyright{usgov}
\acmConference[WPES'20]{19th Workshop on Privacy in the Electronic Society}{November 9, 2020}{Virtual Event, USA}
\acmBooktitle{19th Workshop on Privacy in the Electronic Society (WPES'20), November 9, 2020, Virtual Event, USA}
\acmPrice{15.00}
\acmDOI{10.1145/3411497.3420211}
\acmISBN{978-1-4503-8086-7/20/11}

\newcommand{\etal}{\emph{et al.}\xspace}
\newcommand{\eg}{\emph{e.g.}\xspace}
\newcommand{\ie}{\emph{i.e.}\xspace}
\newcommand{\tabref}[1]{Table~\ref{#1}\xspace}
\newcommand{\figref}[1]{Table~\ref{#1}\xspace}
\newcommand{\secref}[1]{Section~\ref{#1}\xspace}

\usepackage{hyperref}
\begin{document}
\def\UrlFont{\em}
\title{Randomness Concerns When Deploying Differential Privacy}

\author{Simson L. Garfinkel}
\orcid{0000-0003-1294-2831}
\affiliation{%
  \institution{US Census Bureau}
  \streetaddress{4600 Silver Hill Road}
  \city{Suitland}
  \state{MD}
  \postcode{20746}
}
\email{simson.l.garfinkel@census.gov}

\author{Philip Leclerc}
\affiliation{%
  \institution{US Census Bureau}
  \streetaddress{4600 Silver Hill Road}
  \city{Suitland}
  \state{MD}
  \postcode{20746}
}
\email{philip.leclerc@census.gov}

\begin{abstract}
The U.S. Census Bureau is using differential privacy (DP) to
protect confidential respondent data collected for the 2020 Decennial
Census of Population \& Housing.  The Census Bureau's DP system is implemented in the Disclosure Avoidance System
(DAS) and requires a source of random numbers.  We estimate that
the 2020 Census will require roughly 90TB of random bytes
to protect the person and household tables. Although there are critical differences between cryptography and DP, they have similar requirements for randomness. We review the history
of random number generation on deterministic computers\omitt{, including von
Neumann's ``middle-square'' method, Mersenne Twister (MT19937) (the
default NumPy random number generator, which we conclude is
unacceptable for use in production privacy-preserving systems), and the Linux
\emph{/dev/urandom} device}. We also review hardware random number generator
schemes, including the use of so-called ``Lava Lamps'' and the Intel Secure Key RDRAND instruction. We
finally present our plan for generating random bits in the Amazon Web Services (AWS)
environment using AES-CTR-DRBG seeded by mixing bits from \textit{/dev/urandom} and the Intel Secure Key RDSEED instruction, a compromise of our desire to rely on a trusted hardware
implementation, the unease of our external reviewers in trusting a
hardware-only implementation, and the need to generate so many
random bits.
\end{abstract}
\pagestyle{plain}

%
%


\maketitle
\balance
\section{Introduction}

To date, most of the discussion regarding the use of differential
privacy for the US 2020 Census of Population and Housing has focused
on the impact of DP on accuracy and the suitability of DP's privacy guarantee (\eg~\cite{10.1145/3219819.3226070,10.1145/3294052.3322188,NCSL,ruggles1,Hawes2020Implementing}),
and not on the specific details of the Census Bureau's DP implementation.

This paper is divided into two sections. In the remainder of this
section we  present the role of DP in the 2020 Census, discuss
DP's requirements for randomness, contrast DP's requirements for randomness with
those of cryptography, and present related
work. Section \ref{randomness-in-the-2020-das} provides an overview of
the DAS, its randomness requirements, and discusses the engineering
challenges we encountered.

\subsection{DP and the 2020 Census}


As described in the \emph{2020 Census Operational Plan}~\cite{2020op},
the 2020 Census uses data collected from households supplemented
with data from administrative records to create a dataset known as the
Census Unedited File (CUF). This file consists of ``[a]ll person and
household records for the 50 states, D.C., and Puerto Rico.''~\cite[p.9]{2020op}
This file is used to produce the Census Edited File (CEF).
Following the creation of the CEF, the respondent data travels to  a
purpose-built application called the Disclosure Avoidance System
(DAS).

\omitt{ ``DAS is a
subsystem of the Response Processing System, and it applies privacy
controls to microdata as the data are transformed in a flow from the
CEF to the MDF [Microdata Detail File]''~\cite[p.17]{2020op} The MDF,
in turn, is used to produce data products of the 2020 Census.}

The output of the DAS consists of two microdata sets: one containing person records, and a
second in which each record corresponds to a housing unit or group quarters facility. \omitt{The CEF contains a
linkage between each person and the unit in which they reside, but
no such linkage is present in the MDF.}

\emph{Disclosure Avoidance} is a term used by the Census Bureau to
describe techniques employed to limit the risk of a disclosure of respondent
information that would be prohibited by Section 9 of the Census Act
(U.S. Code Title 13), as interpreted by the Census Bureau's Data
Stewardship Executive Policy Committee (DSEP), which is the Census
Bureau's executive policy-setting organ~\cite{disclosure-avoidance-2020-census}.
In 2017, the Census Bureau announced that it would use
DP~\cite{Dwork:2006:CNS:2180286.2180305} as the core
privacy-conferring mechanism for the 2020
Census~\cite{census-csac-2018}.\footnote{An important subtlety is that, though it uses DP subroutines as its core privacy technology, the DAS is not \emph{end-to-end differentially private}, due to the policy requirement that a modest set of ``invariant'' statistics not be altered by the infusion of DP noise. The mathematically provable privacy guarantees conferred by the DAS are weakened by this requirement, but the DAS's privacy guarantee is nevertheless precise and provable, and is similar in form to the guarantees offered by pure DP systems.}

As there was no off-the-shelf mechanism for applying DP to a national census, the
Census Bureau developed its own. Although DP
was created in part with the protection of a national census in mind, the 2020 Decennial Census will be
the first time that a national statistics agency has attempted to
use DP for the purpose that it was created.


Increased transparency of  disclosure avoidance processes
was an important goal in the Census Bureau's adoption of differential
privacy.  In 1990, the Census Bureau adopted a rules-based
``Confidentiality Edit'' termed ``data swapping'' as a privacy
protection mechanism for the original data, and a second technique, called
``Blank and Impute,'' for sample
data~\cite{mckenna2018}. However, no formal proof is available that
these techniques can provide meaningful privacy guarantees against
broad classes of attackers. Nor is it clear that these privacy guarantees are not undermined
by the transparent release of implementation details. Consequently, the Census Bureau
has not released details concerning either the previous disclosure avoidance techniques' implementation,
nor their impact on data accuracy.

By moving to DP, the Census Bureau gained the
ability to provide formal, mathematical proof that meaningful, precise
privacy guarantees hold against broad classes of attackers, and cannot
be undermined by the transparent release of algorithm or
implementation details. This allows the Census Bureau to directly and
explicitly engage with data users and other external audiences, using
publicly available data, concerning the implementation details of the
DAS. This option is particularly useful for helping to explore and
address the trade-offs between privacy loss and accuracy at various
DAS parameter settings.\footnote{Readers familiar with the DP
  literature may note that many DP systems induce error distributions
  which are ``data-independent,'' and can therefore be analyzed even
  without the use of public data sets as a proxy for sensitive
  data. Although the DAS generates a large number of estimators with
  data-independent error distributions, the error distributional
  properties of the MDF are necessarily data-dependent---a side effect
  of the policy requirement that the DAS generate nonnegative
  estimates (which is a single component of the policy requirement
  that the DAS generate microdata, that can be readily manipulated by
  Census systems ``downstream'' from the DAS).}

In the interests of transparency, and to engage external users for preliminary review of the DAS, the Census Bureau
released the source code for the DAS that was used for the 2018 End-to-End Census
Test~\cite{2018-source}, including 62,572 lines of Python source code
and 516 lines of configuration files.


Continuing its engagement with the user community, in October 2019 the
Census Bureau re-released data from the 2010 Census using a
prototype for the 2020 Census DAS system. Called the 2010 Demonstration Data
Products (2010DDP), this system was the subject of a
December 2019 meeting of the Committee on National Statistics, where
attendees compared the statistical accuracy of these data products with
previous data publications based on the 2010 Census~\cite{workshop-on-2020-census-data-products}. The source code
used to prototype the 2010DDP was released the following
month~\cite{2010-ddp-source}. This code base included 33,853 lines of Python programs and 1263 lines of configuration files.

\subsection{DP and Randomness}

This article focuses on the use of randomness in, and randomness
requirements for, the 2020 DAS. We believe this is of general interest
as a reference implementation of a large-scale DP
system.

As traditionally defined, DP is an \emph{information-theoretic}
requirement that a disclosure avoidance system must satisfy, by which
we mean that DP makes no assumptions about computational limitations
of attackers; moreover, analyses of specific DP algorithms are most
often carried out in real-valued arithmetic and assume access to truly
random variables, ignoring the practical subtleties of floating-point
arithmetic and random number generation. The use of pseudo-random number
generators is of special concern in this article: using a
pseudo-random number generator implies that the information-theoretic
privacy-loss budget may be larger than  claimed, as pseudo-random iterates are not independent in the
strict sense required by information-theoretic definitions. To help
overcome this obstacle, computationally aware adaptations of
DP have been developed that acknowledge and
accommodate the use of cryptographically secure pseudo-random number
generators (CSPRNGs) by modeling adversaries as computationally
bounded~\cite{DBLP:conf/crypto/MironovPRV09}. This is a common
assumption when designing practical cryptographic systems. Hence, implementations and interpretations
of differentially private algorithms can in principle be made consistent with the use of a
pseudo-random number generator.

We shall take up this issue in \secref{developing-the-das}.

\subsection{Comparing DP and Cryptography}

DP grew in part from ideas in cryptography, and the application of
such ideas to the formalized study of privacy, so it is not surprising
that there are
many parallels between DP and cryptography:
\begin{itemize}
\item Applications in both fields have well-defined secrets that need
  to be maintained indefinitely.\footnote{Note, however, that in DP, the meaning of ``secret'' is
  more subtle, and it is only the ``usual'' semantic interpretation of
  the privacy guarantee that does not weaken over time, while some
  alternative privacy guarantees smoothly degrade as more external
  knowledge is accumulated. }
\item Both require strong (pseudo-)randomness guarantees, and ready
  access to random numbers.
\item Side-channel leakage is a threat to implementations of both kinds of systems.
\item Failures are hidden: it's hard to distinguish working systems from compromised systems.
\item Both consider security models in which the attacker has
  full access to the system's source code, and, depending on context,
  may consider attackers with arbitrary prior distributions that can
  fully Bayes' update (i.e., who possess unlimited expertise, side
  information, and computational power), or attackers who face
  computational bounds, or whose background knowledge is in some way
  limited.
\item Given the complexity of algorithm design and
  requirements for correct implementations,
  end-users should generally refrain from creating their systems, and
  instead use algorithms and implementations developed and vetted by
  experts.
\end{itemize}

DP and cryptography also have some important differences.
Of particular note, the DP threat model is somewhat different
from common cryptography threat models. A common cryptography threat model involves three
parties: the message sender (``Alice''), the message receiver
(``Bob''), and the eavesdropper (``Eve''). The DP threat model, by contrast, has
just two parties: the message sender and the message receiver, \emph{who is also
the adversary}.\footnote{Although a reviewer of this manuscript noted that
Private Information Retrieval seems to have a similar threat model, in
that there are also just two parties, in PIR the goal is for the
second party to learn \emph{nothing} about the stored information.
DP's goal is more nuanced.}

Indeed, a fundamental insight of the DP literature is that, when guarding against general adversaries,\footnote{That is, without making special assumptions about the knowledge or computational capacity of attackers.} \emph{every}
novel release based on confidential data leaks some information about that
confidential data to the recipient: if too many queries are answered
too accurately on a confidential database, this necessarily reveals
all of the confidential database's contents. This observation is
sometimes called ``the Fundamental Law of Information Recovery.''\footnote{As of this writing, there seem to be several
  different theorems that might qualify for this name: all share the
  property that if too many queries from a given class can be asked by
  an attacker, with some pre-specified bounds on the noise infused
  into the query answers before release, then the attacker will be
  able, with high probability, to reconstruct exactly all bits in the
  underlying database. The theorems differ in the
  structure of queries used by the attacker, and in the convergence
  rate---\ie, the number of queries required to achieve
  reconstruction. As a practical matter, the query classes treated in
  these theorems generally differ from those actually released in
  practice by national statistical agencies. This may be
  little comfort, given recent, concrete demonstrations of
  high-efficacy reconstruction attacks.}

Put another way, a factor complicating DP over traditional
secret-key cryptography is that the information that is intentionally released (or ``leaked'') in a DP system is related to
the information that needs to be kept confidential (though learning one from the other may require specific background knowledge). For example, the ``aggregate statistics'' intentionally released as part of the Decennial Census necessarily leak some information about the values of individual Census responses---indeed, were the DP tabulations released in the Decennial Census not functions of the sensitive
data, they would be useless. Thus, the use of DP is similar to the use of property-preserving
encryption  schemes~\cite{10.1145/2810103.2813651} or functional encryption schemes~\cite{tcc-2011-23520}, such as order-revealing
encryption~\cite{10.1145/2976749.2978379}, in that making the
protected data more useful also inherently makes it more
revealing of the very private information that the technique seeks to protect.

This difference in threat models is natural, as DP's central motivation is quite different from
the traditional application of secret-key cryptography: rather than seeking to restrict the access of unauthorized parties to
confidential information while enabling complete access by authorized parties, the goal of the commonly understood DP semantics\footnote{``Semantics'' is a common term for formal interpretations given of one of the several precise privacy guarantees that can be shown to hold when using a DP disclosure avoidance algorithm.\\
} is instead to \emph{propose a formal definition for what kinds of
attacker inferences qualify as privacy-eroding}, and then to \emph{quantify how
much of this private information is leaked} when using a given algorithm to achieve a targeted
level of accuracy in a statistical publication.

In the use-cases considered by DP, the typical expectation is that the amount of information
an attacker gains from the data release \emph{should} be
non-zero, so that the released tabulations can be useful; the
privacy loss incurred by individuals as a result of the release must also be non-zero. DP allows for the privacy loss to be precisely defined,
quantified, and sharply bounded.


While practical cryptography deployments assume computationally
bounded adversaries, DP research often focuses on
semantic interpretations with a strictly stronger adversary. The
common DP semantic interpretation assumes an attacker that is computationally unbounded with arbitrarily sophisticated algorithms (specifically, they can fully Bayes' update), and has access to arbitrary
``auxiliary knowledge'' (e.g., from external data sets, or from
directly knowing a data subject) for use in making inferences
about the data subject. 

The most common semantic guarantee given in the DP literature is a promised bound on \emph{how much more any fixed attacker can
learn about any fixed property of a data subject from a differentially private publication based on confidential data than would have
been possible had that data subject's information never been collected in
the first place.} As a simple, concrete example, this interpretation of the privacy guarantee justifies statements such as: ``For small $\epsilon$, after a release of data based on a differentially private mechanism, an attacker's confidence that you are of Voting Age cannot be very much larger than it would have been had the same data release been performed but without your data included.''

Put a bit more tersely, we can summarize this by saying that the risk of an attacker learning anything about you (or, in fact, anyone else) is about the same (for small $\epsilon$), regardless of whether you participate in data collection. This upper bound increases smoothly as the
privacy-loss ``budget'' parameter ($\epsilon$) increases\footnote{It is notable that there is no finite $\epsilon$ at which these bounds become meaningless; they always impose some non-degenerate bound on attacker learning. However, they do become loose quite quickly as $\epsilon$ grows, as the bounds depend exponentially on $\epsilon$. In cases where a small $\epsilon$ cannot be justified, alternative semantic statements can be made that apply with smaller bounds, but only by qualitatively weakening the privacy guarantee---for example, we may have to compare an attacker's beliefs relative to a world in which only a portion of a person's data record was not used, rather than their entire record not being used.}.

In secret-key cryptography there is typically only one bound that
is relevant: the attacker should be able to make inferences about an
encrypted message's contents no better than they could have without
seeing the encrypted message in the first place. In this restricted
sense, information-theoretic cryptographic guarantees correspond to
the special case of the guarantee intended by DP systems when
$\epsilon=0$.  This statement comes with two caveats: first, it
ignores the previously discussed differences in attacker
models. Second, standard cryptographic guarantees are
better identified with non-standard DP semantic guarantees, rather
than the usual DP guarantees. DP guarantees typically compare attacker inference after a
release to attacker inference in a world where a data subject did not
participate, whereas cryptographic guarantees are more strongly related to DP semantics
that compare attacker inference after a release to attacker
inference before the release. Guarantees of this kind are more
demanding\footnote{These guarantees essentially involve considering the change before collecting any data, not just if a single person's data had not been collected.} and necessarily
degrade---though smoothly---as more auxiliary knowledge is gathered,\footnote{The reason for this is that auxiliary knowledge can involve learning about probabilistic dependence (or correlation, colloquially) between distinct persons' records, which allows for improved inference about a target person, using information concerning other persons' records.}
unlike the usual DP guarantees (except when $\epsilon=0$) where they
correspond to the usual information-theoretic secrecy promises desired for
encrypted messages.


Because $\epsilon=0$ implies that released tabulations will be useless, deploying and using a DP system
\emph{inherently} involves making and understanding social choices and
economics. Setting the privacy-loss budget ($\epsilon$) fundamentally requires making a
trade-off between the usefulness of the data release and preserving
confidentiality. The data custodian must determine the cost of leakage
and the benefit of tabulation release at a given accuracy. How best to resolve this trade-off is necessarily a \emph{policy} question; algorithm designers can help to provide more \emph{efficient} algorithms, with higher accuracy for a given $\epsilon$, but the ``correct'' choice of $\epsilon$ is not a question that can be resolved through design of better algorithms. The central research question in much of the DP literature is, therefore, whether there are more efficient mechanisms that have more statistical utility for the same $\epsilon$.

%

This paper focuses on a property DP shares with cryptography: the need for large amounts of high-quality random
numbers. For use of DP in large-scale applications,
randomness requirements are driven not by the memory footprint of the
underlying microdata, but by the number and scale of
output tabulations published. In the DAS's case, the randomness required
  is further driven by the need to build intermediate
  ``histograms''---counts of synthetic records of all possible
  types---in order to readily convert DP statistics into microdata,
  and to provide some (typically small) expenditure of privacy-loss
  budget on even arbitrarily complex statistics of the true
  data. Thus, while the data for the Decennial Census can be stored in a
few tens of gigabytes, protecting its output statistics will
require the DAS to use roughly 90TB of random data.

%
%

\subsection{Related Work}\label{related-work}

\subsubsection{Historical Roots}
The history of random numbers is littered with the corpses of methods
that were known not to be truly random when they were deployed but
were incorrectly thought to be good enough for the task at hand.

Tippett published a book of random numbers for use by
computers\footnote{That is, human computers~\cite{10.5555/1536845}.} in 1927~\cite{tippett-1927}.
Following in this tradition, the RAND Corporation published \textit{A Million Random Digits with 100,000 Normal
  Deviates} in 1955\footnote{RAND's book also found uses outside of
  computing:  ``a nuclear submarine commander kept a copy of the
  book with him to chart courses during evasive
  maneuvers.''~\cite{soldiers-of-reason}.\\
  }~\cite{rand1m}. During
production, RAND discovered that the ``electronic roulette wheel'' built for generating
the numbers exhibited statistical bias and required adjustment prior
to publication~\cite{rand1m-online}. Such books of random numbers pose
two practical problems: the sequence is available to anyone who
has a copy of the book, making the numbers unsuitable for security or
privacy applications. And since the book is printed, the numbers are
not easily available for use by electronic computers.

Indeed, early \emph{electronic} computing efforts at Los Alamos required
large quantities of random numbers implement Stanislaw Ulam's ``Monte Carlo
Method''~\cite{mc-beginning}. Early computers did not have an intentional source of
usable randomness, so von Neumann invented the ``middle-square''
method, which generates a stream of digits by
starting with an $n$-digit integer, squaring it, and then extracting the
middle $n$ digits~\cite{jvn-various-techniques}. von~Neumann recognized
that the resulting sequence of digits,  while seemingly randomly, were
entirely predictable. Reflecting on this contradiction, von~Neumann wrote:

\begin{quote}
``Anyone who considers arithmetical methods of producing random digits
is, of course, in a state of sin. For, as has been pointed out several
times, there is no such thing as a random number---there are only
methods to produce random numbers, and a strict arithmetic procedure
of course is not such a method.''~\cite{jvn-various-techniques}
\end{quote}

\subsubsection{Statistical requirements for randomness}

Every modern statistics
package provides facilities for generating seemingly random
values. Typically these numbers are drawn from a specific distribution
such as a uniform distribution, a Gaussian distribution, or some other
named and well-studied distribution. Internally, modern statistics
packages implement a pseudo-random number generator (PRNG) that takes
a single value as a \emph{seed} and emits a sequential series of
numbers, qualitatively similar to von Neumann's method but typically
with a longer period. As with von Neumann's method, if the same seed is provided for two runs,
those two runs will present the same sequence of random values. This
measure of repeatability is useful when developing statistical
programs because it allows for regression testing.

In 1998, Matsumoto \etal introduced the Mersenne Twister, a fast
pseudo-random number generator based on a linear congruential
generator with a period of $2^{19937}-1$. Known as MT19937,
the generator was widely adopted---for example, it was adopted by the
popular Python NumPy numeric library.

In 2007 L'Ecuyer and Simard introduced \emph{TestU01}~\cite{10.1145/1268776.1268777}, a collection of
144 statistical tests for random number generators, mostly gathered
from the  literature, and developed three test suites for
randomness: \emph{SmallCrush}, \emph{Crush} and \emph{BigCrush}. \omitt{They
applied these tests to MT19937; on an AMD Athlon~64 processor the
tests completed with run times of 14 seconds, 1 hour, and 5.5 hours
respectively}. MT19937 failed the linearity tests of \emph{Crush} and
\emph{BigCrush}, which are specifically designed to detect linear
generators.

In July 2019, the Python NumPy scientific library 1.17.0 was
released, with an upgraded random number system that allows the use of
pluggable ``bit generators'' to produce the random distributions. This
allows any source of seemingly random bits to be used for generating a
sequence of seemingly random integers in a given range or floating-point
numbers that match a desired distribution. In this way,
questions of speed, reproducibility, and even suitability for
operation in a parallelized environment are pushed from NumPy down to
the design and implementation of the bit generator. NumPy version 1.19
includes support for four bit generators: MT19937, PCG~64~\cite{pcg64,pcg-wrap},
Philox~\cite{6114424}, and SFC64~\cite{sfc64}.

\subsubsection{Cryptographic requirements for randomness}

Most modern cryptographic systems are based on Kerckhoffs's
principle,  that the design should be public and all of the
security provided by the system should reside in the secrecy of the
cryptographic key. As a result, cryptographic applications require
sources of randomness to create unguessable keys.

The underlying rationale for using Kerckhoff's principle is that strength of
a secure systems depends both on the strength of the algorithm
\emph{and} on the inability of the attacker to determine
the data protection key. Algorithms are hard to
design and, once deployed, they are hard to
replace. Keys, in contrast, should be relatively easy to
change.  So it makes sense for most cryptography users to rely on
algorithms and implementations that have been publicly developed and
vetted, for the simple reason that most users lack the resources to do
as good a job. 

Eastlake \emph{et al.}~\cite{rfc1750,rfc4086} extensively documents the
requirements for randomness in modern computer systems.

Cryptography researchers have spent considerable effort developing and
analyzing random number generators~\cite{rgennaro, Ruhault_2017}. A
common design is to create an \emph{entropy pool}, which is typically
implemented as a buffer into which entropy
from some source of perceived randomness is added through a bit-mixing
function, and from which bits are extracted through the use of a cipher or
cryptographic hash function.\footnote{For readers unfamiliar with an
  the term \emph{entropy pool}, imagine a jar filled with red and blue
  marbles that is periodically stirred by the computer's background
  tasks. When a program wants random numbers, a mechanical arm reaches
  into the jar, causing additional mixing, and emerges with a fistful
  of random bits. These are read in sequence and then returned to the
  jar.}

The American National Standards Institute (ANSI) adopted the ANSI X9.17/X.931 standard for Financial
Institution Key Management (Wholesale) in 1985. The standard presented
a design for a secure pseudo-random number generator that combines
timestamps with a statically keyed block cipher to produce the
pseudo-random output. This design was widely adopted, even though it
obviously requires that the static key be kept secret.

Cohen \etal performed a ``systematic study of publicly available FIPS 140-2 certifications for
hundreds of products that implemented the ANSI X9.31 random number
generator, and found twelve whose certification documents use of
static, hard-coded keys in source code, leaving the implementation
vulnerable to an attacker who can learn this key from the source code
or binary''~\cite{10.1145/3243734.3243756}.
One conclusion of the  study
is that certified implementations of cryptographic systems are not
necessarily secure implementations, and that even professional
programmers make mistakes when it comes to implementing random number
generators, even when they are following a widely used
standard.

Other attacks have been found against the ANSI X9.31 standard
(e.g. ~\cite{6819177}), demonstrating that the mere fact that an
algorithm has been standardized does not imply that it is secure.

Reviewers of the Census Bureau's initial decision to use Intel's Secure Key (ISK) 
as the sole source of randomness cited
the experience with ANSI~X9.31 as an example of why it is important to
build systems that do not have a single point of failure---for
example, by using multiple sources of entropy and by frequently
reseeding the system's CSPRNG.

\subsubsection{Non-determinism in Linux}


Although the EDVAC, the EDSAC, and other computers designed in
von~Neumann's time were intended to be deterministic machines---the
occasional moth no longer being of concern given the field's
transition to electronic tube technology---modern computers have many
sources of non-de\-term\-in\-ism. Weaver \etal identified
``operating system interaction [12], program layout [13], [1],
measurement overhead [14], multi-processor variation [15], and
hardware implementation details [13], [16]''  as potential sources of
non-determinism~\cite{6557172}. (Note: references in the Weaver quotation refer to those in
  Weaver~\cite{6557172}, not the references in this paper.) However, Das \etal suggested that it is
a mistake to use such sources for security purposes, in part because
they are either insufficiently unpredictable, or they are susceptible
to manipulation~\cite{8835366}.

Nevertheless, there has been considerable attention to the use of
such non-determinism as a source of entropy for entropy pools, the most prominent example
probably being the Linux Random Number Generator (LRNG).

M\"uller authored a 196-page report analyzing  the
LRNG in kernels 4.10 through 4.20 and 5.0 through
5.5~\cite{lrng-muller}. M\"uller's assessment states ``[t]he goal of the
assessment is to determine whether the Linux-RNG is able to provide
100 bits, the threshold defined by [TR021021], of entropy early after
a system boot'' and concludes:

\begin{quote}
  ``Applying the general Linux-RNG entropy heuristics, the Linux-RNG
  significantly underestimates the available entropy...This allows the
  conclusion that when the \emph{getrandom} system call unblocks, sufficient
  entropy has been accumulated to be available for use cases with
  strong cryptographic requirements.  The measurements of the
  available entropy during boot for virtual environments and native
  hardware hardly differ. Thus, the conclusion is equally applicable
  to both environments.
\end{quote}
\begin{quote}
    ``It is important to note that this conclusion is only applicable
    to environments with a high-resolution time stamp. Hardware
    architectures with a low-resolution time stamp will not have
    significant amounts of entropy after boot.''~\cite{lrng-muller}
\end{quote}

The Amazon Elastic Map Reduce Kernel used by the Census Bureau is
Linux 4.14.128, and so it is covered by M\"uller's report. The source
code for the version 4.14.128 kernel's random device can be downloaded
from
\url{https://git.kernel.org/pub/scm/linux/kernel/git/stable/linux.git/tree/drivers/char/random.c?h=v4.14.128}.
The \hfill Linux \newline 4.14.128 kernel's random device is \copyright 2017 by Jason
A. Donefeld and is 2343 lines long, of which 658 lines
are comments, addressing some of the concerns previously raised by Gutterman \etal~\cite{1624027}.

The Linux kernel maintains two entropy pools: a larger
pool that receives entropy from the top-half of the kernel using the
\emph{\_mix\_pool\_bytes} function, and a smaller entropy pool
that receives bytes mixed in using the \emph{fast\_mix} function that
is called during system interrupts serviced by the bottom half of the
kernel.\footnote{The ``top-half of the kernel'' refers to the code
  that is invoked by system calls, while the ``bottom-half of the
  kernel'' is the portion that is invoked in response to hardware
  interrupts. Although the terms are widely used and appear in the
  kernel source code, we were unable to
  find a suitable reference to their origin}

 Pseudorandom bytes are extracting by running the CHACHA20 stream cipher~\cite{rfc8439} over a portion of the larger
pool, which the source code claims creates a CRNG (cryptographicaly
strong random number generator). (The Linux version 3 kernel extracts
bytes using the SHA-1 cryptographic hash function; Linux stopped using
SHA-1 in version 4.8.)

We tested the r5.24xlarge AWS Linux VMs used by the Census Bureau and
found that the systems's \emph{gettimeofday()} system call had
microsecond resolution, in that we were able to observe
single microsecond increments in the time returned by the system call
as it was repeatedly called from a tight loop in a C~program.



RedHat Linux provides two device interfaces to the
randomness: \url{/dev/urandom}, which is the output of the ostensible
CSPRNG,\footnote{We say ``ostensible'' not because we doubt whether
  \url{/dev/urandom}'s use of CHACHA20 constitutes a CSPRNG, but
  because we are aware of no formal proof of
  \emph{/dev/urandom}'s security properties.} and
\url{/dev/random}, which maintains a counter of the amount
of entropy that has been added to the entropy pool and blocks if there
is not sufficient entropy remaining until more entropy has been
added.

The \emph{random} device exports four interfaces that are meant to be
called from other parts of the kernel to add entropy:

\begin{description}
    \item[\emph{add\_device\_randomness()}] adds information such as ``MAC addresses or serial numbers, or the output of the RTC (real time clock)'' that are ``likely to differ between two devices.''
    \item[\emph{add\_input\_randomness()}] adds randomness from user input, such as mouse movements or keyboard strikes.
    \item[\emph{add\_interrupt\_randomness()}] adds randomness from the interrupt layer
    \item[\emph{add\_disk\_randomness()}] adds randomness based on the disk seek times.
\end{description}

At this point, the reader may be concerned that a Linux kernel running
in a data center may lack a sufficient source of entropy, at least immediately
following system start-up:
\begin{enumerate}
\item MAC addresses and serial numbers are predictable, as is the
    output of the RTC.
  \item There is no mouse or keyboard on servers in a data center,
    so these are not sources of randomness.
  \item Hardware interrupts during the boot process are predictable.
  \item Disk seek times are predictable on systems equipped with
    solid-state drives, a fact noted in the source code.
\end{enumerate}

The Linux source code acknowledges this
possibility, and provides a small script that
saves 512 random bytes from \emph{/dev/urandom} into the file
\emph{/var/run/random-seed} at system shutdown, and then copies this file
back into \emph{/dev/urandom} at system startup. The source code recommends
that these shell scripts be used to preserve the entropy pool between reboots.
But this approach does not work with Amazon's
Elastic Map Reduce (EMR): since virtual machines are cloned from a master image, there
    is no ability for each VM to have its own, unique entropy pool
    carried in the file \emph{/var/run/random-seed} between system reboots.
This is, in fact, the very situation that M\"uller's report attempts
to address!

Recognizing that the data center needed an improved source of
randomness, Linux added support for CPU-based hardware
random number generation in 2018~\cite{linux-random}, as discussed in \S\ref{controversy}.


\subsubsection{Hardware Random Number Generators}

An alternative approach to using machine randomness is to generate
random numbers using an external entropy source.  Such an approach
based on the movements of the liquid within a Lava Lamp and captured by a digital camera  is described by Noll \etal in US Patent
5,732,138~\cite{5732138}. The Internet hosting company Cloudflare
famously has a wall of Lava Lamps in its office and uses them to seed
the allegedly cryptographically secure pseudo-random number generators that the
company reports using in its Internet serv\-ices~\cite{randomness-101}, and has more
recently developed an entropy service that mixes entropy from five
sources in different countries, using a combination of Lava Lamps,
seismic sources in Chile, environmental noise, extraterrestrial noise,
and other sources~\cite{zd-cloudflare,cf-cloudfalre}.

We
  attempted to evaluate whether Lava Lamps in fact provide sufficient
  randomness for DP after a Census Bureau official
  suggested using them as a source of randomness. 
  
  Lava Lamps were
  invented in 1963 by Edward Walker, and gained popularity in the
  1970s. The lamps use an incandescent light bulb to heat blob of wax
  inside another material. As the wax heats up, it moves upwards
  through the liquid, at which point it cools down and descends. The
  wax blob's movements are governed by thermodynamically controlled
  microcurrents and is often regarded to represent a chaotic system,
  because small variations in energy distribution resulting from
  micro-fluctuations of the power line and in the air surrounding the
  lamp are amplified and result in unpredictable movements of the
  wax. US Patent 5,732,138 conjectures that Lava Lamps are ``chaotic
  systems,'' and states that the patent could be implemented ``for
  example, by taking pictures of a moving freeway, clouds, or lava
  lamps''~\cite{5732138}. However, a conjecture in a patent is not
  scientific validation.  It is also unclear if the patent's invention
  requires that the functioning Lava Lamp be a chaotic system in the formal sense of that phrase, and not merely an
  unpredictable system. In any event, we were informed by a Census Bureau safety officer that Lava Lamps are prohibited from the US Census Bureau's Suitland, MD,
headquarters due to fire-safety concerns. 
Fortunately, thermal noise is
also present at the atomic level, and its use there does not
constitute a fire hazard.

Another approach is to rely
directly on quantum mechanics as a source of randomness.
ID
Quantique introduced such a true random number generator in 2001; the
device has most recently been reduced to a silicon chip for inclusion
in 5G smartphones~\cite{id-quantique}. Los Alamos National Laboratory partnered with Whitewood Security to create
the Entropy Engine~\cite{entropy-engine,lanl-awards}, a hardware random number
that plugs into a PCI Express slot and can generate 350 Mbit/s of true
random numbers.

In January 1999, Intel announced that the forthcoming Pentinum III
microprocessor would include a hardware true random number generator (TRNG) and a
unique processor serial number (PSN) in each chip. The TRNG implementation
sampled thermal noise 32 bits at a time into a shift register
which was then processed with SHA-1. So long as a few bits of the
noise change from moment-to-moment, the output of the SHA-1 function
is thought to be unpredictable. Reviewing Intel's published design, Guttman warned that the generator could fail
without detection~\cite[p.238]{guttman1}.

Before the Pentium III's random number generator could be subject to
further analysis, its PSN was
attacked by privacy activists and, eventually, a legislative panel of
the European Union~\cite{eu-pIII}. Under pressure, Intel removed the PSN from the
``Taulatin'' (130nm) version of the Pentium III and it was not present
in the Pentium IV. The random number generator was on the same section
of the chip as the PSN, resulting in the loss of the TRNG as
well. However, updated versions of both returned (with little fanfare)
in the ``Ivy Bridge'' (22nm) series of Intel's Core processors (Core
i3, i5 and i7)~\cite{intel-2009}. The updated PSN is now termed the
PPIN (Protected Processor Identification Number). The updated random
number generator, code-named Bull Mountain Technology, now
has the official name Intel Secure Key~\cite{intel-drng} (ISK).

\subsubsection{Intel Secure Key}

In this section we review the extensive documentation regarding
Intel's hardware random number generator, and discuss the controversy
surrounding its adoption.

Intel states that the design requirements for a random number
generator (RNG) is that each new value be \emph{statistically
independent}, that the numbers be \emph{uniformly distributed}, and that the
sequence be \emph{unpredictable}, in that ``an attacker cannot guess
some or all of the values in a generated sequence. Predictability may
take the form of \emph{forward} prediction (future values) and
backtracking (past values)''~\cite{intel-drng}.

The Intel ``Digital Random
Number Generator'' (DRNG) is implemented as a module that is separate from the cores on
Intel's multi-core chips. The hardware entropy source is a noisy
asynchronous self-timed circuit that outputs a random stream of bits
at 3 GHz. The hardware entropy source
feeds into a hardware AES-CBC-MAC based ``conditioner'' to ``spread''
the ``entropy sample into a large set of random values.'' It is not a
conventional entropy pool, in that the contents are flushed with
every random draw.

Intel provides two unprivileged user-level instructions for accessing
the DRNG: RDSEED and RDRAND~\cite{rdrand-rdseed}. Both instructions
are available in 16, 32 and 64-bit versions that allow seemingly
random bits to be stored in the designated destination register.

The RDSEED instruction passes the output of the AES-CBC-MAC based
conditioner through a ``non-deterministic random bit generator'' that
is compliant with NIST~SP~800-90B and C~(drafts) (as of November 17,
2012) and provides the bits directly to the
caller~\cite{intel-drng}.
According to Intel,  ``RDSEED is intended for seeding
a software PRNG of arbitrary width.''

The RDRAND instruction causes the output of the AES-CBC-MAC based
conditioner to be used as an input for a AES-256 circuit operating in
CTR mode.
The RDRAND
instruction draws from the output of the AES circuit, with an
upper-bound of 511 128-bit samples being used for each random seed.
Intel states that this is a
``cryptographically secure pseudo-random number generator'' (CSPRNG) that is
compliant with NIST SP~800-90A.\footnote{As with the Linux random
  number generator's use of CHACHA20, the claim that Intel's use of
  AES constitutes a CSPRNG is not, as far as we are aware, supported
  by formal proof. Instead, the security of AES is justified by the
  fact that decades of study has failed to discover significant vulnerabilities in it.}


Intel uses the terms \emph{multiplicative prediction resistance} and
\emph{additive prediction resistance} to describe the difference
in security between the RDSEED and the RDRAND instructions. We could
find no other reference for these terms, so we provide Intel's:

\begin{quote}
\hspace{1pc}``The numbers returned by RDSEED have multiplicative prediction
resistance. If you use two 64-bit samples with multiplicative
prediction resistance to build a 128-bit value, you end up with a
random number with 128 bits of prediction resistance ($2^{128} \times 2^{128} =
2^{256}$). Combine two of those 128-bit values together, and you get a
256-bit number with 256 bits of prediction resistance. You can
continue in this fashion to build a random value of arbitrary width
and the prediction resistance will always scale with it. Because its
values have multiplicative prediction resistance, RDSEED is intended
for seeding other PRNGs.

\hspace{1pc}``In contrast, RDRAND is the output of a 128-bit PRNG that is compliant
to NIST SP 800-90A. It is intended for applications that simply need
high-quality random numbers. The numbers returned by RDRAND have
additive prediction resistance because they are the output of a
pseudo-random number generator. If you put two 64-bit values with
additive prediction resistance together, the prediction resistance of
the resulting value is only 65 bits ($2^{64} + 2^{64} = 2^{65}$). To ensure that
RDRAND values are fully prediction-resistant when combined together to
build larger values you can follow the procedures in the DRNG Software
Implementation Guide on generating seed values from RDRAND, but it's
generally best and simplest to just use RDSEED for PRNG seeding.''~\cite{rdrand-rdseed}
\end{quote}

\omitt{
Essentially, RDRAND only reseeds periodically, so knowledge of its internal state
  between reseeds could be used to predict future states until the
  reseed happens. RDSEED in a sense reseeds on every request, so
  knowing the first 64 bits doesn't help with the second. Intel states
  that RDSEED should be used if the random draw is used ``to seed
  another pseudo-random number generator,'' whereas RDSEED should be
  used ``for all other purposes.''~\cite{rdrand-rdseed}}

The DRNG monitors its output using Online Health Tests (OHTs) and
Built-In Self Tests (BISTs) and shuts the system down if the output of
the DRNG fails to meet statistical quality tests. As of 2014, the
system could produce a maximum of 800 MB/sec of random data, an upper
bound for all threads and all cores on the CPU. Because of this limit,
if a random value is not available, the RDRAND and RDSEED instructions
set the CPU carry flag (CF) if the returned value is actually random;
otherwise CF is set to zero and the destination register is
cleared. Intel recommends that applications calling RDRAND attempt 10
retries when in a tight loop if either instruction returns and CF is
not set. For RDSEED, Intel recommends that a PAUSE instruction be
inserted in the retry loop, and a maximum of 100 retries be
performed. It is not clear what a program should do when the retry
limit is exceeded.

Intel notes that RDSEED is not available on all processors, and that
it can be simulated on such processors that have RDRAND by using RDRAND to generate 512 128-bit samples and cryptographically mixing the results to assure that one of the values was a fresh value from the DRNG and not the
result of AES counter mode.

\subsubsection{ISK Adoption and Controversy}\label{controversy}

In 2012, Cryptography Research (a consulting firm that was highly respected in the cryptography community) conducted an independent review of
the DRNG and concluded ``the Ivy Bridge RNG is a robust design with a large margin of safety that ensures good random data is generated even if the ES [Entropy Source] is not operating as well as predicted.''~\cite{cryptography-research}

In 2015, Shrimpton and Terashima presented ``A Provable-Se\-cur\-ity
Analysis of Intel's Secure Key RNG'' at
EUROCRYPT~\cite{eurocrypt-2015-27241} and concluded that the security
guarantees offered by Intel Secure Key ``tell a mixed story:''

\begin{quote}
``We find that ISK-RNG lacks backward-security altogether, and that
  the forward-security bound for the ``truly random'' bits fetched by
  the RDSEED instruction is potentially worrisome. On the other hand,
  we are able to prove stronger forward-security bounds for the
  pseudorandom bits fetched by the RDRAND
  instruction.''~\cite{eurocrypt-2015-27241}
  \end{quote}

ISK generated considerable
controversy in the Linux community.
Hardware random number generators would seem  ideal  for
providing a source of randomness in a data center, especially if one
boots virtual machine snapshots as is the case with Amazon Web
Services.
For this reason, the Linux kernel includes the functions
\url{arch_get_random_seed_long()} and \url{arch_get_random_long()} to
return hardware-gen\-e\-ra\-t\-ed random numbers. On Intel-based systems,
these functions gateway to the RDRAND instruction, and the output of that
instruction is added to the Linux entropy pool.

Nevertheless, by 2013 there was already broad knowledge of ISK
within the Linux community and a growing desire on the part of some
developers to use it, while a reluctance on the part of others to do
so. After all, ISK could not be readily audited, because it was
implemented in silicon. In 2013, a petition on \emph{change.org}
requested that Linux remove the use of the RDRAND instruction from the
\url{/dev/random} device. The fear was that a hardware backdoor might
give Intel, and perhaps other organizations, the ability
to predict its random output.

The concern over ISK is similar to concerns that were raised regarding the
adoption of the Dual Elliptic Curve Deterministic Random Bit Generator
(Dual\_EC\_DRBG) in NIST SP800-90, ``Recommendation for Random Number
Generation Using Deterministic Random Bit Generators
(Revised)''~\cite{sp800-90}. The Dual\_EC\_DRBG algorithm depends on several pre-specified
constants, and the way that those constants were created can make the
algorithm vulnerable to attack. The summer that Dual\_EC\_DRBG was proposed, concerns were raised that
there might be a ``secret backdoor'' in the standard~\cite{schneier-wired}.
In 2013, those concerns about Dual\_EC\_DRBG  were
confirmed~\cite{nyt-rng}. NIST responded by issuing guidance stating ``NIST strongly
recommends that, pending the resolution of the security concerns and
the re-issuance of SP 800-90A, the Dual\_EC\_DRBG, as specified in the
January 2012 version of SP 800-90A, no longer be
used.''~\cite{nist-guidance}\footnote{To avoid this sort of problem,
  other algorithms that require constants sometimes rely on a
  so-called ``nothing-up-my-sleeve number,'' in which the number is
  chosen from a well-known mathematical sequence, such as the decimal expansion of $\pi$. For
  example, the SHA-2 Secure Hash Algorithm uses the square roots and
  cube roots of small primes for its constants.}
A 2015 article by the Director of Research at the National Security
Agency described the agency's ``failure to drop support for the
Dual\_EC\_DRBG'' after vulnerabilities were identified in 2007 as ``regrettable.''~\cite{nsa-regrets}.

Linux's inventor Linus Torvalds refused to remove support for RDRAND from the
kernel, because RDRAND's entropy is mixed into the Linux
entropy pool, and not used to replace the Linux entropy
pool. Torvalds responded to the \emph{change.org} petition:

\begin{quote}
  ``Short answer: we actually know what we are doing. You don't.
  ~\\[1.5ex]
  ``Long answer: we use rdrand as
\emph{one} of many inputs into the random pool, and we use it as a way
to \emph{improve} that random pool. So even if rdrand were to be
back-doored by the NSA, our use of rdrand actually improves the
quality of the random numbers you get from /dev/random~\cite{linus-response}.
\end{quote}

Torvalds's comments were supported by the comments of Linux engineer
Theodore Y. Ts'o, who noted: ``I am so glad I resisted pressure from
engineers working at Intel to let \url{/dev/random} in Linux rely
blindly on the output of the RDRAND infrastructure. Relying solely on
an implementation sealed inside a chip and which is impossible to
audit is a BAD idea.''~\cite{tso-hacker-news}.

Nevertheless, the petition's request \textit{was} ultimately
implemented in the Linux kernel. In August 2018, the Linux v1.19-rc1
release candidate kernel included a flag that allowed the kernel to be
compiled without support for RDRAND~\cite{register-2018,linux-random}.
(We note that even if kernel support for RDRAND is disabled, the
instruction can still be accessed from user-level programs, since use
of the instruction does not require privilege.) And in 2019, the Linux
kernel added ``sanity checking'' to the output of the hardware random
number generator, after the discovery that the hardware random number
on some AMD-based systems stopped providing random values after a
suspend/resume cycle~\cite{phoronix-rdrand,kernel-tip}.

In June 2020, Ragab \etal published an attack called CrossTalk that
allows one user-level process running on an Intel-based computer to
eavesdrop on the output of the RDRAND instruction run in another
process by observing the Intel CPU's shared ``staging''
buffer~\cite{ragab_crosstalk_2021}. This attack is not relevant to the DAS, since
we assume that no unauthorized software is running in the Census
Bureau's secure computing environment. However, some systems running
ISK have now been patched to address this issue. After the patch is applied, RDRAND reportedly generates random values with only 3\% of
its unpatched performance.

\subsubsection{Use of RDRAND in Statistical Software}

The Python programming language adopted MT19937 as its default random
number generator in Python~2.3 and still uses it for version~3.8, the
3.9.0b1 release candidate, and the Python~3.10 development tree. The
documentation notes ``Warning: The pseudo-random generators of this
module should not be used for security purposes. For security or
cryptographic uses, see the secrets module.''~\cite{python3.8-random}
The Python secrets module includes a function \emph{SystemRandom} which
calls the Python function \url{os.urandom()} as a source of randomness; on
Linux systems this reads from \url{/dev/urandom}, which (as noted
above) may incorporate entropy mixed-in from RDRAND. (Note: because it
can block and will supply only a small number of random values, \emph{/dev/random} may not be appropriate for use in production
statistical software.)

As noted above, the Python numpy numeric package now provides for user-supplied random bit generators. Thus, it is now straightforward to combine modern versions of Numpy
with an RDRAND-based bit-generator such as Sheppard's
\emph{randomgen}~\cite{randomgen}. This is a non-standard mode of
operation, and requires that \emph{randomgen} be separately
downloaded.
We analyzed \emph{randomgen.RDRAND}'s behavior and source
code. We verified that NumPy run in this configuration is in fact using \emph{randomgen.RDRAND}
by running on an 2011 MacBook Air laptop and observing that the NumPy
random number generator raised an exception, as the MacBook Air's processor lacked the ISK. 
However, our analysis of  the
\emph{randomgen.RDRAND} source code revealed that, as of July 9, 2020,
it did \emph{not} check the Carry Flag (CF) as recommended in Intel's
software implementation guide~\cite{randomgen-246}. This
implementation error was reported to Sheppard and it was promptly corrected.

We also analyzed the source code for the Python Package Index (pypi) rdrand model
version 1.5.0~\cite{rdrand-150} and found that it did properly
implement the CF check.

\subsubsection{Randomness Requirements for DP}

As noted above, DP's requirements for the quality of randomness
sources are similar to the corresponding requirements in cryptography. 


Because pure DP's definition is stated information theoretically, pure
DP is inconsistent with the the use of a PRNG, even a
CSPRNG. Mironov \etal introduced several computationally-aware DP variants, in which attackers are assumed to face clearly defined computational bounds (and so can no longer fully Bayes' update); these attacker restrictions are similar to those used in the definition of a CSPRNG. ``The good news is that a DP mechanism coupled with a
  [cryptographically secure] PRNG will satisfy the stronger definition
  of the two. This is the theoretical underpinning for conveniently
  ignoring the issue of (information-theoretic) DP vs computational
  DP''~\cite{DBLP:conf/crypto/MironovPRV09}.

Dodis \etal considered the impact of an imperfect randomness
source on the privacy guarantees offered by DP by
comparing them to the privacy guarantees associated with using
such a randomness source to generate cryptographic
keys~\cite{10.1007/978-3-642-32009-5_29}. The authors also discuss the
impact of using ``infinite-precision'' mechanisms that rely on an
infinitely long random tape in $\{0,1\}^*$ and discuss how to approximate
it with a tape that offers randomness of finite precision. However, as
the randomness  sources modeled in Dodis \etal are incomparable to CSPRNGs,
  they are not directly relevant to use of CSPRNGs in the DAS.


The impact of mathematical precision on the privacy guarantees of
DP was taken up by Mironov's 2012
paper~\cite{mironov2012on}, which presented an attack that allowed the
compromise of underlying confidential data due to ``irregularities of
floating-point implementations of the privacy-preserving Laplacian
mechanism.'' The attack is effective because, in some settings, the differences between IEEE floating-point representations and arithmetic can cause the least significant bits of certain queries to leak much more information about individuals in the confidential database than the information-theoretic definition of DP implies---i.e., these differences cause the actually achieved $\epsilon$ to be much larger than the $\epsilon$ claimed on the basis of interpreting floating-point implementations of probability distributions as equivalent to their real-valued descriptions. This theme is
further explored by Gazeau~\emph{et al.}~\cite{GAZEAU201692}.



\section{Randomness in the 2020 DAS}\label{randomness-in-the-2020-das}

This section provides an overview of the 2020 DAS, discusses its
requirements for randomness, and then discusses how those requirements are achieved.

\subsection{Overview of the DAS}\label{overview-of-the-das}

The 2020 DAS is a Python-Spark
(pyspark) application deployed on an Amazon Elastic Map Reduce (EMR)
cloud-computing cluster. EMR runs on top of Amazon Linux, with
nodes being configured to run either the pyspark driver program, or as
pyspark workers. Currently DAS is running on EMR version 5.25.

The \emph{Amazon EMR: Amazon EMR Release Guide}~\cite{amazon-emr-release-guide} details all of the version
numbers of the open source Apache software used by each EMR release,
but it does not mention the version of Amazon Linux on which the
release is based. Although it is possible to run EMR with a custom
Amazon Machine Image (AMI)~\cite{restricted-use-microdata}, the Census
Bureau is using a standard AMI, specifically Amazon Linux
AMI release 2018.03, which identifies itself as ``amzn'' and is
based on RedHat Fedora.


\subsection{Operation of the DAS}
The operation of the 2020 DAS has been described by
the Census Bureau in other publications, including an overview
presented to the Census Bureau's Scientific Advisory
Committee~\cite{census-csac-2018}, the published design specification
for the DAS~\cite{census-2010-design-specification}, and a
draft academic article describing the so-called TopDown Algorithm (TDA)~\cite{census-top-down-2010}.

Briefly, the algorithm runs twice to produce privacy protected microdata: once, to produce the table containing person-level
microdata, and a second time to produce the table containing microdata
for housing units and group quarters facilities. For each table, a histogram of counts is computed at
each geographic level for the United States that is described by the
Census Bureau's geographical ``spine'' (currently the US as a whole,
the states and D.C., the counties and county
equivalents, the census tracts, and the blocks). Each of the cells of
each of these histograms, as well as a set of queries chosen to reflect the to-be-published Census tabulations, is then protected using an existing
DP mechanism ``using the Laplace
mechanism~\cite{10.1007/11681878_14}, Geometric
Mechanism~\cite{10.1145/1536414.1536464}, or more advanced techniques
such as the high dimensional matrix
mech\-anism~\cite{10.14778/3231751.3231769}.''~\cite{census-top-down-2010} These protected values are called the ``noisy measurements.''

Next, the algorithm performs two optimizations on the US-level
histogram\footnote{Data for Puerto Rico is processed using a separate
  pass of the same algorithm.} subject to external knowledge
constraints (e.g., that state-level population totals must not be
noisily perturbed): the first optimization minimizes squared error to
the DP estimates of the histogram cell and query values, while
requiring that the output be a non-negative, floating-point-valued
estimate of the true histogram.\footnote{That is, this optimization imposes non-negativity
  and \emph{self-consistency} as requirements: after this optimization
  is complete, all queries on the microdata can be calculated based
  off of it, and the estimates attained will be unique. This circumstance does not hold for the initial DP measurements,
  where we may, for example, have an estimated total population that is not the
  sum of the estimated white alone and not-white-alone
  populations.\\
  } The second optimization pass forces all counts
to be integers, performing a variant of controlled rounding while
maintaining external knowledge constraints.

Following the US-level optimization, the \-algorithm performs
a\-nalogous optimizations in which the counts in the US-level histogram are allocated to the state-level histograms, while minimizing
differences between the counts assigned to the states and the counts computed from the noisy measurements within each state. Once again, this
is a two-step optimization process. This process then repeats for every geographic node in each geographic level of the geographical hierarchy until all counts are distributed to individual Census blocks. Finally, each block's histogram is expanded into the microdata that the histogram specifies. This process is referred to as
\emph{post-processing} in some Census Bureau
presentations.\footnote{This use of the term post-processing is a
  reference to the technical use of the same term in the DP
  literature, where the ``post-processing property'' refers to
  a straightforward but important result: that the DP guarantee cannot
  be undermined by performing further processing on the output of a DP
  algorithm, so long as the confidential data is not directly accessed
  in doing so.} These optimizations provably do not violate the
guarantee provided by TDA's use of differentially private mechanisms, although the use of invariants in this post-processing does, as we previously commented, imply that the achieved privacy guarantee is qualitatively weaker than it would be under pure DP.

The application of DP is a relatively small but essential part of the TDA: the DP subroutines in use by the DAS tend to be much simpler and faster than those used by the DAS for optimization-based post-processing. However, it is essential for the DP mechanism to be correct, as these are the principal source of the DAS's privacy guarantees. Among other considerations, this means that acquiring suitable high-quality randomness is necessary for the DAS to achieve its purpose.


\subsection{Randomness Requirements for the DAS}

We estimate the randomness requirements for the DAS by observing
that the privacy mechanism starts by computing a histogram of counts
for every geographical unit at every geographical level. (In
  practice, each ``run'' of the DAS actually consists of two distinct
  runs: one for the persons table and one for the housing unit/group
  quarters facility table. For each of these runs the DAS usually
  manipulates two histograms simultaneously---a principal histogram
  concerned with variables that heavily interact with one another in
  published tabulations, and a much smaller histogram featuring a
  small set of variables that mostly do not interact with those in the
  main histogram. For simplicity we ignore this much smaller histogram
  in our descriptions and calculations here. Additionally, our
  calculations focus on the histograms rather than individual queries,
  as the histograms dominate the randomness requirements by orders of magnitude, and the queries can change from run to run, depending on configuration file specifications.) We compute the total number of bits for both runs of the top-down algorithm for the United States (but not for Puerto Rico, which is run separately) in \figref{requirements} and find it to be a minimum of 90TB of random data. \figref{requirements} modestly underestimates the randomness requirements because the final DP workload includes not just each histogram cell, but additional queries of summary statistics (some of which have not yet been determined).

\begin{figure*}
\begin{minipage}{\textwidth}
  \begin{align}
\textrm{Total \# Random Bits}
  & = 64 \times ( \textrm{Total \# protected histogram cells}) \nonumber \\
  & = 64 \times ( |H_p| + |H_u|) (\# geounits) \nonumber \\
  & = 64 \times ( |H_p| + |H_u|) \sum{G_\textit{geolevels}} \nonumber \\
  & = 64 \times ( |H_p| + |H_u|) (G_\textrm{nat} + G_\textrm{state equivs} + G_\textrm{county equivs} + G_\textrm{tracts} + G_\textrm{block groups} + G_\textrm{blocks} ) \nonumber \\
  & = 64 \times ( (42\times2\times116\times2\times63) + (2\times9\times2\times7\times4\times2\times522)) (1 + 51 + \textrm{3,143} + \textrm{73,782} + \textrm{217,550} + (\sim \textrm{8,000,000} )) \nonumber \\
  & \approx 7 \times 10^{14} \textrm{bits} \nonumber \hspace{1em}(\approx 90\textrm{TB})\\
\textit{Where} \hspace{2pc}|H_p| & = \textrm{Size of the person-level histogram} \nonumber \\
|H_u| & = \textrm{Size of the unit-level histogram} \nonumber \\
G_{\textrm{level}}  & = \textrm{The number of geounits at geolevel \textit{level}} \nonumber
 \end{align}
\end{minipage}
\caption{Randomness requirements for the US run of the 2020 DAS,
  including 50 states and the District of Columbia. Geography counts
  for county equivalent and tracts taken from 2010 Census. It is
  estimated that there will be 8 million habitual blocks for the 2020 census.}\label{requirements}
\end{figure*}

\subsection{Threat Models}
In deciding upon the source of randomness for the DAS, the development
team engaged in several threat modeling exercises. We assumed that an
attacker would have access to the entire DAS software and hardware stack, including
the actual implementation of the TDA used to generate the published
statistics, the Python runtime environment, the Linux operating
system, the same hardware on which the DAS had been run, and detailed
information of the network configuration. We assumed that the attacker
would have all data publications produced from the confidential data,
and that the attacker could combine these publications with
significant external knowledge. For example, we assumed that an
attacker was likely to know the rough age, race and sex
distributions of every community in the U.S., since the Census Bureau
already publishes this information as part of the American Community
Survey. We also assumed that the attacker has unbounded mathematical skills and computational capabilities, although we did assume that the attacker could not crack AES-256.

Except as otherwise noted, however, we exclude most
risks pertaining to the correctness of software and hardware
implementations. That is, we assume that attackers do not have access
to the systems on which the TDA is running, as this would give the
attacker access to the underlying confidential data, rendering moot
DP's privacy protections. We likewise assume that Linux kernel on
which the TDA executes matches the source code for the Linux kernel
that we reviewed.

\subsection{MT19937 and DP}\label{mt19937-concerns}

Because MT19937 is the default PRNG in Python and was the default
Numpy RNG prior to Numpy 1.17, MT19937 has been widely used in
DP demonstrations. Indeed, MT19937 was used in the
initial prototype implementation of the DAS.

It is inadvisable to use  MT19937 in a production system that is intended to protect confidential data: though MT19937
has a large period of $2^{19937}-1$, MT19937 has substantial security vulnerabilities.

MT19937 maintains its internal state as a vector
of 624 32-bit unsigned integers, and its output can, as a byproduct of this, be predicted after
observing just 624 output iterations. Indeed, there is publicly
available source code that implements this attack
(\eg~\cite{kmyk-predict}), and there are  several
user-friendly blog posts outlining how to perform a variant of this attack~\cite{mtattack-1,mtattack-2,mtattack-3}.

In an application like the DAS, which requires protecting large,
sparse histograms, the vast majority of the protected histogram counts are the result of taking true counts of zero and adding noise derived from sequential draws of the RNG. Thus, if an attacker knows that certain populations are not present in a geographic area, the attacker can immediately infer the noise iterates from viewing the DP query estimates. This situation is problematic if MT19937 is used, as it reduces the security of the implementation to the difficulty of inverting Laplace or Geometric distribution random draws: if these can be inverted, the attack on MT19937 can be carried out, and the DP guarantees unravel entirely for  geographic units where the necessary auxiliary knowledge is available.

For the 2020 Census, the Census Bureau is in fact considering the release of
the so-called ``noisy measurements,'' or the
raw values, of each cell in each histogram after the noise has been
added, as these enjoy in principle the same privacy guarantees as the final microdata. The histograms being considered for the DAS are currently
217,550 and approximately 8 million cells.
Since the vast majority of
these cells are likely to be zero, if MT19937 is used, an attack of this kind may be quite practical. Here is a sketch:

\begin{itemize}
  \item This attack relies on finding a run of
    313 cells known to correspond to true counts of zero, and being
    able to invert the noise algorithm such that the output of the
    MT19937 algorithm can be determined. With these output values, the internal state of the MT19937 engine is then determined.
  \item The internal state is validated by running the MT19937 algorithm two steps forward to determine
    the next 64 output bits. The 64 output bits predict the contents of the 313\textsuperscript{th} cell. If the 313\textsuperscript{th} cell matches the prediction, then the internal state of the MT19937 algorithm is validated.
  \item Once the state is validated, the amount of noise that was added can be inferred for every other cell in the histogram, and the privacy protection mechanism is undone.
\end{itemize}

This attack requires being able to transform a 64-bit noise
value into two successive 32-bit draws, which may require some additional thought and computation to achieve, so this attack is not immediately practical. Nevertheless, it is not desirable for the security of a DP implementation to rely on the difficulty of inverting noise-distribution sampling functions, which are not designed with security in mind. Thus, it is clear that MT19937 is inappropriate for production DP applications. We believe that only CSPRNGs should be used for production privacy applications.



\subsection{Developing the DAS}\label{developing-the-das}
During initial efforts to develop the DAS, 
the development team discussed whether or not it would be desirable to have a
``repeatable'' source of random numbers to support regression
tests~\cite{10.1145/3267323.3268949}. Initial efforts to create a
repeatable sequence failed: the research team discovered
that Apache Spark's scheduler was non-de\-ter\-min\-is\-tic, and that the same
workload might be scheduled simultaneously on multiple nodes if there
were available resources. Although there was discussion that the random seed could be made dependent upon the geographical unit, ultimately it was decided that this was unnecessary; instead, approximate repeatability could be achieved by caching the DP measurements and building functionality to re-load from them, as this functionality would be required for other purposes regardless.\footnote{Caching the DP measurements in fact ensures exact repeatability, except for a small degree of non-auditible non-determinism that can occur in the Gurobi Optimizer (e.g., when using it in a mode where multiple optimization algorithms are deployed in parallel, terminating when the first such algorithm terminates), on which the DAS relies.}

The DAS development team turned its attention to the Python random
number generator in the early part of 2018. Learning of the problems
with MT19937, but unwilling to create its own implementation of the
Laplace or exponential distribution random samplers, the initial work-around employed was to reseed the MT19937 generator with a read of \url{/dev/urandom} on every scalar draw. This proved to be unacceptably slow as the scale of problems the DAS needed to address increased beyond what was required for early testing.

The DAS development team was familiar with the Intel RDRAND
instruction and learned that Intel had developed its own Python
distribution which made use of hardware acceleration on the Intel
platform. The team assumed that the Intel Python distribution adopted
RDRAND for use in the Numpy RNG, but it had
not.

Intel also created \url{mkl_random}, ``a
NumPy-based Python interface to Intel (R) MKL [Math Kernel Library] Random Number Generation
functionality''~\cite{mkl_random,MKL}.  This software was installed on
the DAS development clusters and used for the 2018 Decennial Census End-to-end test.
However, additional testing revealed several implementation flaws,\footnote{See
  \url{https://github.com/IntelPython/mkl_random/issues}} causing the
  2020 DAS development team to stop using the software.

\omitt{
The DAS geometric mechanism is implemented by instances of \emph{class
  GeometricMechanism(DPMechanism)}, which are initialized with the
parameters \emph{epsilon}, \emph{sensitivity}, and
\emph{true\_answer}, which is an array of confidential values.
The random number generator is instantiated from \emph{self.rng}, which is
initialized as \emph{MKLRandom()}. This class
applies a two-tailed geometric distribution to a numpy.ndarray using a straightforward construction of the two-tailed geometric distribution as a difference of two one-tailed geometric distributions. There are two instance methods: \emph{pmf()} (probability mass function) and
\emph{inverseCDF()}.
}

\omitt{
The file \emph{primitives.py} also
  contains two legacy mechanisms. The first is
  contained in \emph{class LaplaceMechanism(DPMechanism)}, which takes the same
  arguments as \emph{GeometricMechanism} and an additional argument
  \emph{prng}. The Laplace mechanism is implemented
  by calling \emph{prng.laplace()}. The \emph{prng} argument is an
  object that is a subclass of
  \emph{np.random.RandomState}.
\vspace{1pt}
The second legacy mechanism
  defined in the file is
  \emph{class GaussianzCDPMechanism()}, for which the class
    constructor takes the arguments \emph{rho},
    \emph{l2\_sensitivity}, \emph{true\_answer} and \emph{prng}.
}

\subsection{Randomness and the 2020 DAS}

Although M\"uller would seem to be the final word on the subject, the
DAS development team had concerns about using \emph{/dev/urandom}
as the sole source of randomness for the privacy mechanism of the 2020
Census. Our first concern was that it was surprisingly difficult to
verify that the version of \emph{/dev/urandom} that M\"uller had
reviewed was the same version that the DAS team was running, as there
is no obvious mechanism for verify the integrity of modules that make
up a kernel running in the AWS environment. Our second concern
validating whether or not the entropy seed of the AWS device might be
inadvertently replicated as part of the cluster boot process. Our
third concern was performance: there is only a single \emph{/dev/urandom}
device and corresponding entropy pool for each AWS kernel, creating a
single-threaded bottleneck in our otherwise parallelized
implementation. With 96 cores per cluster, such a bottleneck might
adversely impact performance.

The alternative to using \emph{/dev/urandom} is to implement a
user-level CSPRNG. Such a CSPRNG could be seeded from
\emph{/dev/urandom} or from a suitable source of hardware entropy.

Two sources of hardware entropy in the AWS environment are the ISK (on appropriately equipped systems) and a network-ac\-ces\-si\-ble hardware security module for
generating cryptographically strong random numbers that Amazon
provides~\cite{aws:kms:GenerateRandom}. Amazon's service is part of the
AWS Key Management Service (KMS), which is an integrated system for
managing and using both symmetric and asymmetric keys and encryption
algorithms. We tested the KMS GenerateRandom and determined that it
generates 1024 bytes at a time, and has a round-trip-time of
approximately 0.5 seconds per invocation, for a maximum
single-threaded, non-pipelined performance of 2KiB/sec. Although we
could find no documented concerns regarding the operation of the KMS and have no
reason to not to trust the quality of its entropy, we could also find
no reason to trust it other than an appeal to Amazon's authority.

Intel's  documentation states that the DRNG can produce random
data at the rate of 800~MB/sec per CPU chip. Current CPU
chips have faster clock rates and more cores, but it does not appear that the DRNG has been improved significantly or that it is no
longer shared between cores. Assuming that today's chips have the same
DRNG unit and the CrossTalk patches have not been applied, a single
Intel chip could produce the required randomness to protect the 2020
Census using the Census Bureau's DP mechanism in $90
\times 10^{12} \div 800 \times 10^{6} \approx 112,000 $ seconds, or 31
hours.

Of course, the Census Bureau is \emph{not} running the TDA on a single
CPU. Currently, the TDA runs on AWS r5.24xlarge virtual
machines. These systems report 96 cores each arranged in 8 12-core
Intel Xeon chips. For an EMR cluster with 20 workers,
there will be 160~DRNGs, allowing the required randomness to be
computed in about 700 seconds, or about 12~minutes.
(Sheppard observed that  faster times might be achieved
with the same level of security by using the RDSEED
  instruction to seed a software random number generator based on AES
  run in counter mode---essentially a software version of RDRAND
  without the periodic reseeding~\cite{numpy-issue-9365}.)

Given the extensive analysis that the ISK has received, and given the simplicity
of accessing it with a single user-level machine instruction, the DAS
team originally planned to simply use RDRAND as the source of \emph{all} randomness for
the TDA. This approach was discouraged by outside reviewers, who
stressed that a silicon-only implementation could not be
audited. Some reviewers suggested that the Census Bureau rely on the
Linux \emph{/dev/urandom} device exclusively as the source of randomness,
arguing that it already mixes in entropy from RDRAND, as well as from
other sources such as hardware interrupt timing.

The Census Bureau performed a speed test of \emph{/dev/urandom} on the
AWS r5.24xlarge server in AWS GovCloud US-West region and found that a
single-threaded \emph{dd} process could retrieve pseudo-ran\-dom bytes
at the rate of roughly 200~MB/sec, or one fourth the rate of
RDRAND. However, the \emph{/dev/urandom} device is effectively
single-threaded, as there is only one entropy pool and it is protected
with a lock. Four concurrent \emph{dd} processes on the same server retrieved bytes
from \emph{/dev/urandom} with data transfer rates between 52~MB/sec and 54~MB/sec each, the
slight performance boost likely coming from parallelization of the
user-level \emph{dd} code.

On a 96-core machine with 8 CPUs, the 8 ISK devices are able
provide a maximum throughput of 6400~MB/sec---32 times the bandwidth
of random data than can be provided by the \emph{/dev/urandom}
device. On an EMR cluster with 20 worker nodes, this raises the amount
of time required to produce the randomness to 8 hours, which was deemed to be
unacceptable.

\omitt{
(Note that the
system could be made to run twice as fast by requesting from Amazon 40
r5.12xlarge systems, since that would result in there being two
kernels on each physical machine, and thus two entropy pools and two
distinct \emph{/dev/urandom} devices. Because of the complexity of
such an approach, the difficulty in assuring isolation between
multiple kernels on the same VM, and the concern that non-Census
users might make use of another VM running on the same hardware if the
Census Bureau requested r5.12xlarge VMs, the Census Bureau is not
pursuing this option.)}

The Census Bureau's current solution is operate a user-level CSPRNG
based on AES-CTR-DRBG that will be seeded from
user-level calls to RDSEED mixed with output from \emph{/dev/urandom}.
The output of this CSPRNG
will used as a bit generator for NumPy. Although this effectively runs
96 CSPRNGs per r5.24xlarge system, resulting in a significant
performance boost, higher rates of reseeding will lower performance
to the maximum rate at which RDSEED can provide such seeds. Thus, the performance of this
arrangement is tunable, and those tunings have not yet been decided.

The DAS development team has determined that AES side channel and timing attacks are
not of concern in this circumstance, since all software running on the virtual machine is
trusted, the use of r5.24xlarge VMs assures that there are no other
tenants on the physical hardware, and precise timing information will
not be available to potential attackers. Nevertheless, the DAS will
use an AES implementation that employs the native AES acceleration
instructions provided by Intel's microprocessors to prevent side channel attacks.

%

\section{Conclusion}
The need to generate a large number of high-quality random numbers is
a largely unrecognized requirement of a production differential
privacy system. Many DP tutorials and texts assume the
availability of high-quality floating point random numbers taken from
the Laplace, geometric, or exponential distribution. In practice, these examples use MT19937 or PCG64. These algorithms are not CSPRNGs and should 
not be used to protect confidential information. Because this important detail may not be 
obvious to developers looking for a DP implementation, we recommend
that DP tutorials and texts discuss this issue and use CSPRNGs as their randomness sources.

The prototype implementation of the 2020 DAS used MT19937 as seeded by
\emph{/dev/urandom} on a cluster of 96-core Linux servers. After the
DAS development team learned that MT19937 is not secure, the
team changed the DP primitives to use Intel's
RDRAND instruction as the source of randomness, as accessed through the Python \emph{mkl\_random}
library. To avoid relying on the ISK as a single source of
randomness, and after discovering that \emph{mkl\_random} is based on
the closed-source Intel Math Kernel Library, the DAS team pivoted
 to using a user-level random bit generator based on AES-256 and
seeded by the Linux \emph{/dev/urandom} mixed with bits from the RDSEED instruction. Throughout the process, the DAS team found it necessary to review multiple randomness implementations down to the level of assembler code, and found several software quality issues and implementation errors, some of which are discussed in this paper.



The Census Bureau is now four years into the process of modernizing
its disclosure avoidance systems to incorporate formal privacy
protection techniques. This process has proven to be challenging
across disciplines. Beyond the 2020 Census, the Census Bureau intends
to use DP or related formal privacy systems to
protect all of its future statistical publications.

\begin{acks}
At the US Census Bureau, we wish to thank John Abowd for supporting our work
on this project and Donald E Badrak II for his technical
input. We had invaluable assistance and technical input from Galois Inc.'s audit
team (Jose Calderon, Josh Heumann, Scott Moore and Marc Rosen) and Kevin Sheppard (Oxford).
We also had useful input from Paul Bartholomew, Andrew Hong
and Drew Lipman at MITRE, Jim Hughes (UC~Santa Cruz), John
M. Kelsey (NIST), and Sebastiano Vigna (Universit\`a delgi Studi di
Milano). Additional helpful comments were provided by Abraham D. Flaxman (University of Washington), Rich Salz (Akami), and Ted Ts'o
(Google). Our summer intern Joshua
Schmidt provided valuable editorial assistance. We also wish the thank
the anonymous reviewers, who provided useful guidance.

\textbf{\noindent\small The views in this papers are those of the authors, and
  do not represent those of the US Census Bureau.}
\end{acks}
\clearpage
\bibliographystyle{ACM-Reference-Format}
\balance
\bibliography{main,rfcs}

\end{document}